\documentclass[aps,preprint,prb,tighten,showpacs,graphicx,amsmath]{revtex4}

\usepackage[dvips]{graphicx}
\newcommand{\etac}{\eta_c'}
\newcommand{\etap}{\eta_p'}

\newcommand{\Rc}{R_c'}
\newcommand{\Rp}{R_p'}
\newcommand{\q}{q'}
\newcommand{\mypicwidth}{0.8\columnwidth}

\newcommand{\EPL}{{\it Europhys.~Lett.~}}
\newcommand{\PRL}{{\it Phys.~Rev.~Lett.~}}
\newcommand{\PR}{{\it Phys.~Rev.~}}
\newcommand{\JCP}{{\it J.~Chem.~Phys.~}}

\newcommand{\JPCM}{{\it J.~Phys.: Condens.~Matter~}}
\newcommand{\MP}{{\it Mol.~Phys.~}}
\newcommand{\JCIS}{{\it J.~Coll.~Int.~Sci.~}}

\begin{document}

 
\title{Mixtures of Charged Colloid and Neutral Polymer: 
Influence of Electrostatic Interactions on Demixing and Interfacial Tension}
\author{Alan R. Denton\footnote{Electronic address:
{\tt alan.denton@ndsu.edu}}}
\affiliation{Department of Physics, North Dakota State University,
Fargo, ND, 58105-5566}
\author{Matthias Schmidt\footnote{Electronic address:
{\tt mschmidt@thphy.uni-duesseldorf.de}}}
\affiliation{Institut f\"ur Theoretische Physik II, 
Heinrich-Heine-Universit\"at D\"usseldorf, Universit\"atsstra\ss e 1, 
D-40225 D\"usseldorf, Germany}

\date{\today}
\begin{abstract}
The equilibrium phase behavior of a binary mixture of charged colloids
and neutral, non-adsorbing polymers is studied within free-volume theory.
A model mixture of charged hard-sphere macroions and ideal, coarse-grained, 
effective-sphere polymers is mapped first onto a binary hard-sphere mixture 
with non-additive diameters and then onto an effective Asakura-Oosawa model 
\break [S. Asakura and F. Oosawa, \JCP {\bf 22}, 1255 (1954)].
The effective model is defined by a single dimensionless parameter --
the ratio of the polymer diameter to the effective colloid diameter.
For high salt-to-counterion concentration ratios, a free-volume approximation 
for the free energy is used to compute the fluid phase diagram, which 
describes demixing into colloid-rich (liquid) and colloid-poor (vapor) phases.
Increasing the range of electrostatic interactions shifts the demixing
binodal toward higher polymer concentration, stabilizing the mixture.  
The enhanced stability is attributed to a weakening of polymer 
depletion-induced attraction between electrostatically repelling macroions.  
Comparison with predictions of density-functional theory reveals a 
corresponding increase in the liquid-vapor interfacial tension.
The predicted trends in phase stability are consistent with observed 
behavior of protein-polysaccharide mixtures in food colloids.
\end{abstract}

\pacs{61.20.Gy, 64.70.Ja, 82.70.Dd, 05.20.Jj, 05.70.-a}

\maketitle




\section{Introduction} 

Mixtures of colloidal particles and free (non-adsorbing) polymer coils
dispersed in a solvent are among the most intensively studied soft matter 
systems~\cite{pusey91,poon02,tuinier03review,aarts03swet,aarts04capw,
aarts04codef}.
The conceptual analogy between colloids and atoms, similarities in 
thermodynamic phase behavior between colloidal suspensions and atomic 
systems, and the relative ease of tuning polymer-induced effective colloidal 
interactions, make colloid-polymer mixtures valuable model systems for 
probing connections between microscopic interparticle interactions and 
macroscopic properties in a variety of materials.
The observed phases, distinguished by composition and structural order,
include most of the equilibrium phases familiar in simple molecular systems, 
{\it e.g.}, vapor, liquid, crystal, as well as nonequilibrium states, 
such as glasses and gels~\cite{pusey91,note1}.
Complementing their fundamental importance, colloid-polymer mixtures 
have diverse industrial applications, {\it e.g.}, to coatings, petroleum
products, pharmaceuticals, and many foods, where polymer additives are 
used to control phase stability and rheological properties.

Addition of free polymer can substantially modify effective interactions 
between colloidal particles through the mechanism of entropic depletion.
When two colloidal surfaces approach to a separation closer than the 
typical diameter of a polymer coil, the entropic cost to the coil of 
distorting its average spherical conformation tends to exclude the polymer.
The resulting depletion of polymer from the space between the colloids 
creates an imbalance in polymer osmotic pressure that can induce 
effective attractions between colloids.  Attractions of sufficient range 
and strength, depending on the relative size and concentration of polymer, 
can drive bulk demixing into colloid-rich (liquid) and colloid-poor 
(vapor) phases.

The first, and conceptually simplest, statistical mechanical model to 
qualitatively describe the polymer depletion mechanism and the associated 
phenomenon of depletion-driven demixing was the model of Asakura and 
Oosawa~\cite{asakura54}, developed independently by Vrij~\cite{vrij76}.  
The Asakura-Oosawa (AO) model regards the colloidal particles as 
hard spheres and the polymer coils as effective ideal spheres -- 
mutually noninteracting, but interacting with the colloids as hard spheres.  
Neglect of polymer-polymer interactions can be reasonably justified 
for polymers in a theta solvent.  The phase diagram 
corresponding to the AO model has been determined by a variety of methods, 
including thermodynamic perturbation theory~\cite{gast83}, 
free-volume theory~\cite{lekkerkerker92,aarts02}, 
density-functional theory~\cite{schmidt00cip}, 
and Monte Carlo simulation~\cite{dijkstra99,bolhuis02phasediag,vink05pre}. 
Recent effort has also been devoted to calculating the liquid-vapor
interfacial tension of phase-separated colloid-polymer 
mixtures~\cite{vrij97,brader00,brader02,louis02tension,moncho-jorda03,
aarts04cahn,vink04jcp,vink04codef,moncho-jorda05,vink05}.

Although certain topologically constrained polymers ({\it e.g.}, stars 
and microgels) are close to spherical in shape, linear-chain polymers 
are random walks whose shapes are spherical only on average. 
To explore the significance of nonspherical conformations, 
more explicit segmented-chain polymer models have been studied via
Monte Carlo simulation~\cite{meijer-frenkel,dickman94,louis00} 
and integral-equation theory~\cite{fuchs-schweizer}.
Effective polymer-polymer interactions have been modeled by combining 
Monte Carlo and integral-equation methods within a ``polymers as 
soft colloids" framework~\cite{louis00}.
In recent work, we examined fluid-fluid demixing within several 
variations of the classic AO model, incorporating a third component 
(hard needles)~\cite{schmidt02cpn}, polymer-solvent 
interactions~\cite{schmidt02cpps},
polymer-polymer interactions~\cite{schmidt03cintp}, and 
colloid-induced polymer compression~\cite{denton02cpoly}.
The latter study incorporated intrinsic polydispersity in the polymer 
radius of gyration.  
Influences on phase behavior of colloid polydispersity~\cite{fasolo04} 
and of polymer chain length polydispersity~\cite{sear97polydisperse,fasolo05}
have also been studied theoretically.

While neutral colloid-polymer mixtures have been widely investigated,
much less attention has been devoted to mixtures of charged species. 
In the case of charged colloids or charged polymers (polyelectrolytes), 
electrostatic repulsions between macroions, screened by counterions 
and salt ions in solution, compete with depletion-induced attractions 
and can modify phase behavior.  
In pioneering experiments~\cite{sperry84,gast86}, phase separation 
was observed in mixtures of charged colloids and neutral polymers.
Gast {\it et al.}~\cite{gast86} interpreted their observations by 
applying thermodynamic perturbation theory to an effective one-component 
model, incorporating electrostatic repulsion and depletion-induced 
attraction into an effective pair potential between colloids.  
Subsequent measurements of force profiles in mixtures of charge-stabilized 
colloidal oil-in-water emulsion droplets and ionic surfactant 
micelles~\cite{mondain-monval95} and in mixtures of colloidal particles
and charged macromolecules~\cite{walz} directly demonstrated 
the potential for electrostatic interactions to modify depletion forces.  
More recently, liquid-vapor separation and gelation were observed in 
mixtures of charged colloids and ionic wormlike micelles~\cite{petekidis02} 
and depletion potentials induced by charged rods were measured in 
colloidal rod-sphere mixtures~\cite{helden04}.
In a recent theoretical study, liquid-state integral-equation methods 
were used to model the structure and phase behavior of mixtures of 
charged colloids and polyelectrolytes~\cite{ferreira00}.

The purpose of the present paper is to propose, for a simple model of a 
charged-colloid -- neutral-polymer mixture, an alternative theoretical approach
that, in contrast to Ref.~\onlinecite{gast86}, treats the two components 
on an equal footing.  The theory is based on mapping the binary mixture 
onto an effective AO model, governed only by excluded volume interactions,
and applying free-volume theory~\cite{lekkerkerker92} and classical 
density-functional theory~\cite{schmidt00cip}.
Within this conceptual approach, we explore the qualitative influence of 
colloidal charge on demixing and find that increasing the range of 
electrostatic interactions can significantly stabilize the mixture against 
demixing and correspondingly increase the liquid-vapor interfacial tension.

The remainder of the paper is organized as follows.  The model system is 
defined in Sec.~\ref{SECmodel}.  The mapping onto the effective AO model and 
the free-volume theory are described in Sec.~\ref{SECtheory}.  Results for 
the fluid-fluid demixing phase diagram and interfacial tension are presented 
and discussed in Sec.~\ref{SECresults}.
Finally, Sec.~\ref{SECconclusions} closes with a summary and conclusions.

\section{Model}
\label{SECmodel}
The system of interest comprises charged colloidal particles, their 
dissociated counterions, and free polymer coils, all dispersed in 
an electrolyte solvent.
The multi-component mixture of macroions, microions (counterions and 
salt ions), polymers, and solvent molecules spans a range of length 
and time scales, presenting severe challenges to explicit modeling approaches.  
To reduce the system to a tractable model, we make several simplifying, 
yet realistic, assumptions.  (1) The colloids are modeled as monodisperse 
charged hard spheres -- a close approximation to many real synthetic 
suspensions.  (2) The microions are modeled as monovalent point charges 
and subsumed into effective electrostatic interactions between macroions, 
as described below.  
(3) The polymer coils are represented, as in the coarse-grained AO model, 
as effective spheres.  Nonspherical conformations, disfavored by lower
conformational entropy, can be reasonably neglected when the polymer coils 
are comparable or smaller in size than the colloids (colloid limit).  
(4) Finally, the solvent is treated as a dielectric continuum, characterized 
by a dielectric constant (primitive model of electrolytes~\cite{Hansen86}).  
An experimental system that closely resembles our model system would be 
charge-stabilized synthetic polystyrene or silica microspheres dispersed 
in an aqueous electrolyte together with a non-adsorbing, nonionic, 
water-soluble polymer, such as hydroxyethylcellulose 
(HEC)~\cite{sperry84,gast86} or polyethylene oxide (PEO).

Interactions between colloidal particles include steric and electrostatic 
repulsions and van der Waals attractions.  Here we assume hard-sphere 
steric interactions and ignore van der Waals interactions, which is valid 
for particles index-matched to the solvent.  Electrostatic interactions 
between colloidal macroions result from bare Coulomb repulsion and screening 
by surrounding microions.  The classic theory of Derjaguin, Landau, Verwey, 
and Overbeek (DLVO)~\cite{derjaguin41-verwey48} predicts that, in a dilute 
suspension, two colloidal macroions at center-to-center separation $r$ 
interact via a pair potential of screened-Coulomb (Yukawa) form:
\begin{equation}
v_{cc}(r)~=~\left\{
\begin{array} {l@{\quad\quad}l}
\infty, & r<2R_c \\
\frac{\displaystyle Z^2e^2}{\displaystyle \epsilon}
\left(\frac{\displaystyle e^{\kappa R_c}}{\displaystyle 1+\kappa R_c}\right)^2
\frac{\displaystyle e^{-\kappa r}}{\displaystyle r}, & r\ge 2R_c,
\end{array} \right. \label{EQvcc}
\end{equation}
where $Z$ is the effective macroion valence, $e$ the proton charge, 
$\epsilon$ the solvent dielectric constant, $\kappa$ the Debye screening 
constant (inverse screening length), and $R_c$ the colloid radius.

More recent statistical mechanical approaches proceed by formally 
integrating out from the partition function the microion degrees of freedom 
to map the macroion-microion mixture onto an effective one-component system 
governed by effective interactions~\cite{silbert91,vanroij97,graf98,
vanroij99,denton_lrt,warren00}, determined by the distribution of microions 
around the macroions.  The microion distribution depends on the response --
in general nonlinear -- of the microions to the macroion charge density.
For sufficiently weakly charged macroions, the microion response can be 
approximated as linear.  A further neglect of microion correlations, 
which is reasonable for monovalent counterions, then recovers the DLVO form 
of effective pair potential [Eq.~(\ref{EQvcc})] with a density-dependent 
screening constant of the form 
\begin{equation}
\kappa=\sqrt{\frac{4\pi(Z\rho_c+2\rho_s)e^2}{\epsilon k_{\rm B}T}},
\label{kappa}
\end{equation}
where $\rho_c$ and $\rho_s$ are, respectively, the number densities of 
colloidal macroions and salt ion pairs, $k_{\rm B}$ is Boltzmann's constant
and $T$ is the absolute temperature.  The nonzero size of the microions 
may modify the microion distribution and thus screening effect, but should 
not qualitatively change the general repulsive form of the effective pair 
potential assumed here.  In practice, since most water-soluble polymers 
are less polarizable than water, $\epsilon$ may be best interpreted as an 
effective dielectric constant of the polymer-solvent mixture and $\kappa$ 
as an effective screening constant.
For the purposes of this paper, the essential point is that the range of
electrostatic repulsion, governed by $\kappa$, can be widely tuned by 
adjusting the ionic strength (salt concentration) of the electrolyte solvent.

As a natural byproduct of the one-component mapping, the total free energy 
contains, in addition to a pair-interaction contribution, a one-body volume 
energy~\cite{silbert91,vanroij97,graf98,vanroij99,denton_lrt,warren00} $E$,
given by
\begin{equation}
\frac{E}{Vk_{\rm B}T}=(Z\rho_c+\rho_s)\left\{\ln[(Z\rho_c+\rho_s)\Lambda^3]
-1\right\}+\rho_s\left\{\ln(\rho_s\Lambda^3)-1\right\}-\frac{Z^2\rho_c
\kappa\lambda_{\rm B}}{2(1+\kappa R_c)}-\frac{Z^2\rho_c^2}{2(Z\rho_c+2\rho_s)},
\label{evol}
\end{equation}
where $V$ is the total volume of the system, $\Lambda$ is the thermal 
wavelength of the microions and $\lambda_{\rm B}=e^2/(\epsilon k_{\rm B}T)$ 
is the Bjerrum length.
In a straightforward physical interpretation, the first two terms in 
Eq.~(\ref{evol}) account for the microion entropy, the third term is the 
interaction of a macroion with its own cloud of counterions, and the final 
term results from charge neutrality.
Because the volume energy depends nontrivially on colloid density, it must,
in general, be included in the free energy.  At relatively high salt 
concentrations ($\rho_s\gg Z\rho_c$), however, the counterions are dominated
by salt ions and the volume energy can be neglected.
We restrict our considerations to this parameter regime.

Within the coarse-grained spherical polymer model, colloid-polymer 
interactions are described by a simple excluded-volume pair potential: 
\begin{equation}
v_{cp}(r)~=~\left\{
\begin{array} {l@{\quad\quad}l}
\infty, & r<R_c+R_p \\
0, & r\ge R_c+R_p,
\end{array} \right. \label{EQvcp}
\end{equation}
where $r$ is now the macroion-polymer center-to-center distance and $R_p$ 
is the polymer radius of gyration.  For simplicity, we assume the solvent 
to be near its theta temperature for the polymer~\cite{deGennes79}, allowing
the polymer to be reasonably modeled as ideal (mutually non-interacting),
{\it i.e.}, $v_{pp}(r)=0$, for all $r$.  In practice, the strength of
polymer-polymer interactions will depend on the properties of specific
polymers and solvents.  

Within the above assumptions, the model system is now completely 
characterized by the colloid-colloid interaction [Eqs.~(\ref{EQvcc}) 
and (\ref{kappa})] and the polymer-to-colloid size ratio $q=R_p/R_c$.  
The thermodynamic states of the system are specified by the 
bulk number densities $\rho_i$, or equivalently the volume fractions, 
$\eta_i=(4\pi/3)\rho_i R_i^3$, of species $i=c,p$.

\section{Theory}
\label{SECtheory}
\subsection{Mapping onto the Asakura-Oosawa Model}
To explore thermodynamic properties of the model colloid-polymer mixture, 
we seek an approximation for the free energy of the system.  We proceed
by first constructing a mapping onto a simpler model system.  A similar 
approach has been developed recently by Tuinier~\cite{tuinier04}.  
Assuming the screened-Coulomb effective pair potential between colloids 
[Eq.~(\ref{EQvcc})] to be relatively steeply repulsive ($\kappa R_c>1$), 
the corresponding contribution to the free energy may be mapped, with 
reasonable accuracy, onto the free energy of an effective hard-sphere 
system~\cite{Hansen86,andersen71} interacting via an effective hard-sphere 
pair potential,
\begin{equation}
v'_{cc}(r)~=~\left\{
\begin{array} {l@{\quad\quad}l}
\infty, & r<2\Rc \\
0, & r\ge 2\Rc,
\end{array} \right. \label{EQvcceff}
\end{equation}
where $\Rc$ is the effective hard-sphere colloid radius.  
A reasonable first estimate of the effective radius is obtained from a
simple thermal criterion,
\begin{equation}
v_{cc}(r=2\Rc)~=~\frac{Z^2e^2}{\epsilon}\left(\frac{e^{\kappa R_c}}
{1+\kappa R_c}\right)^2\frac{e^{-2\kappa\Rc}}{2\Rc}~=~k_{\rm B}T,
\label{thermal-criterion}
\end{equation}
according to which colloids tend not to approach closer than a distance 
at which their interaction energy is comparable to the typical thermal energy.

Assuming high salt-to-counterion concentration ratios ($\rho_s\gg Z\rho_c$), 
the screening constant $\kappa$, and thus $v_{cc}(r)$ and $\Rc$, are
practically independent of density.  In this same limit, the volume energy 
[Eq.~(\ref{evol})] contributes to the free energy density a term that is
only linear in colloid density and therefore irrelevant for phase behavior.
To demonstrate that in this high-salt-concentration regime the effective 
colloid radius may still significantly exceed the bare radius, consider the 
typical example of particles of bare radius $R_c=50$ nm and effective valence 
$Z=500$ suspended at volume fraction $\eta_c=0.01$ in a 1 mM aqueous salt 
solution at room temperature.  In this case, $\rho_s\gg Z\rho_c = 0.016$ mM, 
yet the effective colloid radius is estimated 
[from Eq.~(\ref{thermal-criterion})] to be $\Rc\simeq 67$ nm, {\it i.e.}, 
about 30\% greater than the bare radius, a difference that also well exceeds 
typical colloid polydispersities.

\begin{figure}
\hspace{-2cm}
\includegraphics[width=\mypicwidth,angle=0]{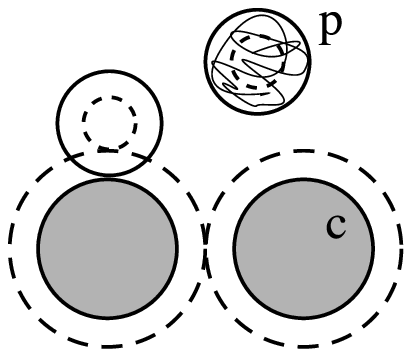}
\caption{\label{FIGmodel}Sketch of the two models of charged colloids 
(c) and (effective spherical) neutral polymer (p).  The solid curves 
indicate the actual sizes of the particles (radii $R_c$ and $R_p$) 
in the original model, and the dashed curves the rescaled sizes 
(radii $R_c'$ and $R_p'$) in the effective Asakura-Oosawa model.  
Note that the depletion layer thickness is the same in both models.}
\end{figure}

The original system has thus far been mapped onto an effective binary 
hard-sphere mixture, governed by the pair interactions of Eqs.~(\ref{EQvcp}) 
and (\ref{EQvcceff}).  As a measure of the relative range of electrostatic 
and excluded-volume repulsions, it is convenient to define the electrostatic 
colloid size ratio, $\xi=\Rc/R_c\ge 1$, as the ratio of the effective and 
bare colloid radii.  The model system is now competely specified by two 
dimensionless parameters, $\xi$ and $q$.  Since the polymer species is ideal, 
this model resembles the AO model, with the colloid diameter merely rescaled.  
However, because colloids and polymers repel at a range of $(R_c+R_p)$, 
rather than $(\Rc+R_p)$, the hard-sphere diameters are no longer additive 
and the rescaling is nontrivial.  Nevertheless, as Fig.~\ref{FIGmodel} 
illustrates, this non-additive mixture can be further mapped exactly onto 
an effective additive AO model by introducing a fictitious effective polymer 
radius $\Rp$, defined via
\begin{equation}
\Rc + \Rp = R_c+R_p
\label{EQradiirelation}
\end{equation}
so as to yield the correct colloid-polymer excluded-volume interaction 
[Eq.~(\ref{EQvcp})].

\begin{figure}
\includegraphics[width=0.75\columnwidth,angle=0]{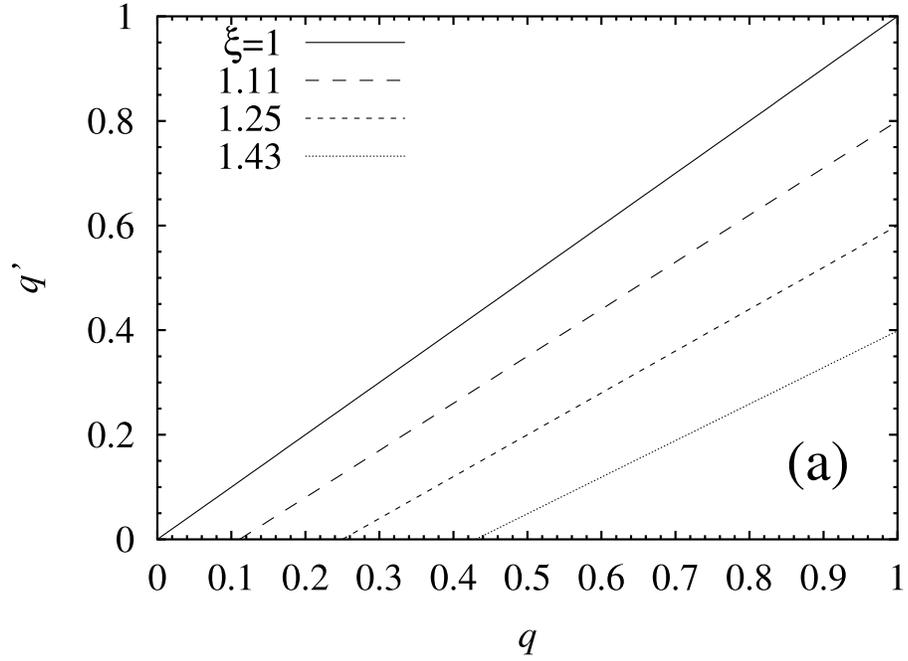} \\[5ex]
\includegraphics[width=0.75\columnwidth,angle=0]{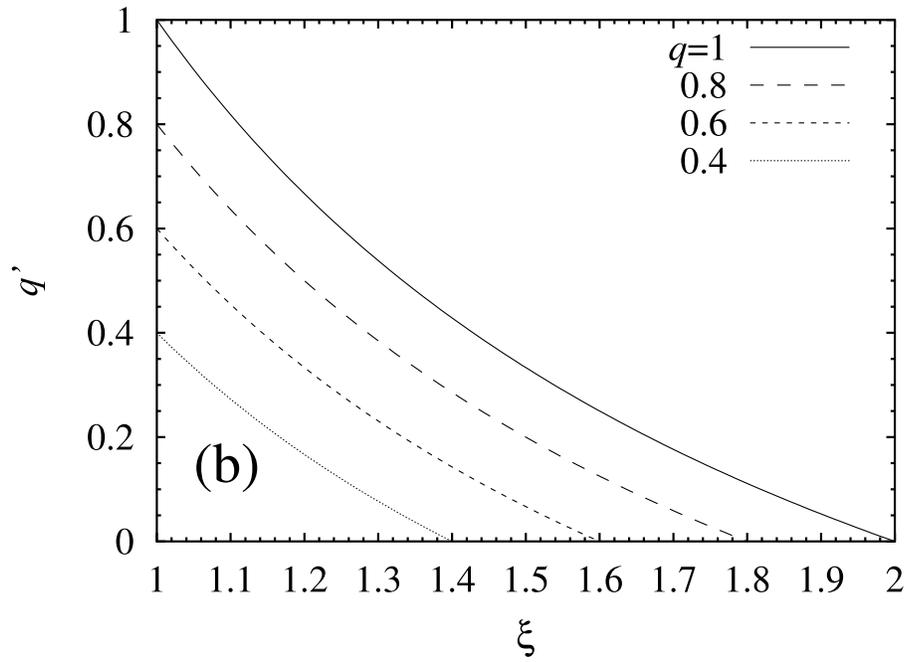}
\caption{\label{FIGbarq}Effective polymer-to-colloid size ratio 
$\q=\Rp/\Rc$ vs.  (a) actual polymer-to-colloid size ratio $q=R_p/R_c$ and
(b) electrostatic colloid size ratio $\xi=\Rc/R_c$.}
\end{figure}

The final effective AO model is characterized by a {\em single} dimensionless 
parameter, namely the effective polymer-to-colloid size ratio 
\begin{equation}
\q = \Rp/\Rc = (q-\xi+1)/\xi, \qquad {\rm for}~\xi\leq 1+q.
\label{qeff}
\end{equation}
As shown in Fig.~\ref{FIGbarq}, the rescaled polymer-to-colloid size ratio
is always smaller than the true ratio ($\q<q$) and with increasing 
electrostatic colloid size ratio, $\q$ decreases, {\it i.e.}, 
the polymers grow effectively smaller relative to the colloids.
For $\xi>1+q$, there is no polymer depletion, and so then $\q=0$.
The thermodynamic states are in turn specified by effective volume fractions 
\begin{equation}
\etac = \frac{4\pi}{3}\rho_c\Rc^3 = \xi^3 \eta_c 
\label{etaceff}
\end{equation}
and
\begin{equation}
\etap = \frac{4\pi}{3}\rho_p\Rp^3 
= [1-(\xi-1)/q]^3\eta_p,
\label{etapeff}
\end{equation}
where we have used the relation $\Rp/R_p=1-(\xi-1)/q<1$,
obtained from Eq.~(\ref{EQradiirelation}).

\subsection{Free-Volume Theory}
Having mapped the original mixture of charged colloids and neutral polymers 
onto an effective AO model, we now consider the Helmholtz free energy,
from which all equilibrium thermodynamic properties may be determined.
The total free energy, $F=F_{\rm id}+F_{\rm ex}$, separates conveniently 
into an ideal-gas term $F_{\rm id}$, which is independent of interactions,
and an excess term $F_{\rm ex}$, which depends entirely on interactions.
For a homogeneous fluid, the ideal-gas free energy density is given exactly by
\begin{equation}
\beta F_{\rm id}/V = \rho_c[\ln(\rho_c\Lambda_c^3)-1] + 
\rho_p[\ln(\rho_p\Lambda_p^3)-1],
\label{EQfid}
\end{equation}
where $\beta=1/k_{\rm B}T$ and $\Lambda_c$ and $\Lambda_p$ are the 
thermal wavelengths of the colloid and polymer species.  
For the excess free energy, we apply the mean-field free-volume theory 
of Lekkerkerker {\it et al.}~\cite{lekkerkerker92}, which predicts 
fluid-fluid phase separation for the original AO model in good agreement 
with simulation~\cite{dijkstra99,bolhuis02phasediag,vink04jcp,vink04codef}.
Within this approach, the fluid excess free energy density is approximated by 
\begin{equation} 
\beta F_{\rm ex}/V = \beta \phi_{\rm HS}(\etac) - \rho_p \ln \alpha(\etac,\q),
\end{equation} 
where $\phi_{\rm HS}(\etac)$ is the excess free energy density of the 
effective one-component hard-sphere fluid and $\alpha(\etac,\q)$ is 
the polymer free-volume fraction, {\it i.e.}, the fraction of the total 
volume accessible to the polymer centers (not excluded by the colloids).
The free-volume fraction is related to the excess chemical potential 
of the polymer via Widom's particle insertion method~\cite{widom63}:
\begin{equation}
\mu_{p,\rm ex}=-k_{\rm B}T\ln\langle\exp(-\Delta U/k_{\rm B}T)\rangle
=-k_{\rm B}T\ln\alpha,
\label{EQmupex}
\end{equation}
where the angular brackets denote an ensemble average over colloid 
configurations and polymer positions, and $\Delta U$ is the change in 
potential energy upon insertion of a polymer -- here infinite if the 
polymer overlaps a colloid and zero otherwise.
An approximation for $\alpha$ is obtained by noting that $\mu_{p,\rm ex}$
also equals the reversible work required to produce a spherical cavity, 
of radius $\Rp$, in a fluid of hard sheres of radius $\Rc$.  
The scaled-particle approximation~\cite{reiss59,lebowitz65} for 
$\mu_{p,\rm ex}$ then yields
\begin{eqnarray}
\alpha(\etac,\q)&=&(1-\etac) \exp(- Ax - Bx^2 - Cx^3),
\label{EQalpha}
\end{eqnarray}
with $x=\etac/(1-\etac)$, $A=\q^3+3\q^2+3\q$, 
$B=3\q^3+9\q^2/2$, and $C=3\q^3$.
Within the same approximation, the excess free energy of the colloids
is given by
\begin{equation}
\beta \phi_{\rm HS}(\etac) = \frac{3 \etac
[3 \etac (2-\etac) - 2(1-\etac)^2\ln(1-\etac)]}
{8 \pi \Rc^3 (1-\etac)^2}.
\label{EQfhs}
\end{equation}
The pressure resulting from Eq.~(\ref{EQfhs}) coincides with the 
compressibility equation of state 
following from the exact solution of the Percus-Yevick integral equation 
for hard spheres~\cite{wertheim63,thiele63}.
Taken together, Eqs.~(\ref{EQfid})-(\ref{EQfhs}) provide an approximation 
for the fluid free energy, from which the fluid-fluid demixing phase diagram 
can be computed.

\section{Results and Discussion}
\label{SECresults}
\begin{figure}
\hspace{-2cm}
\includegraphics[width=\mypicwidth,angle=-90]{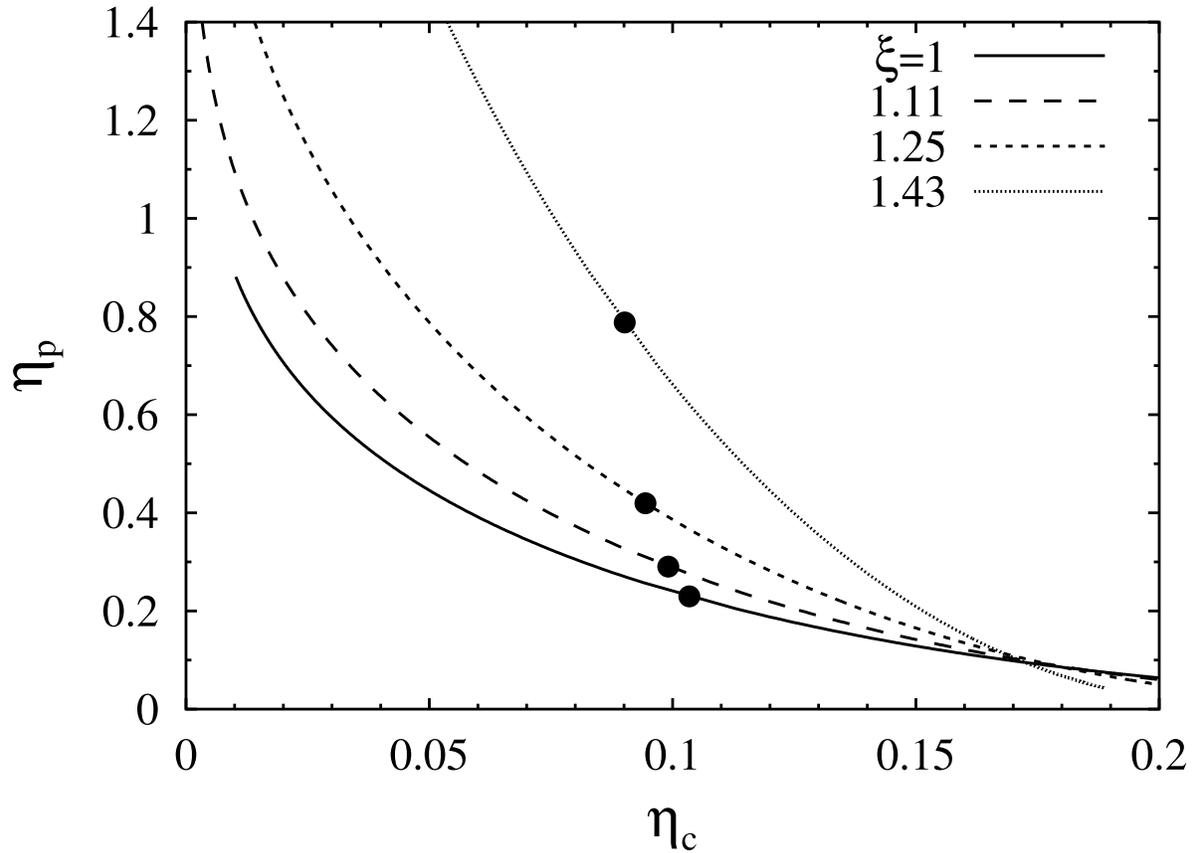}
\caption{\label{FIGpd}Fluid-fluid demixing binodal of charged-colloid -- 
neutral-polymer mixtures as a function of colloid and polymer volume fractions,
$\eta_c$ and $\eta_p$, for fixed polymer-to-colloid size ratio, 
$q=R_p/R_c=1$, and varying electrostatic colloid size ratio $\xi=\Rc/R_c$.  
From bottom to top, $\xi=1, 1.11, 1.25, 1.43$, 
corresponding to increasing effective colloid radius.
Symbols represent the respective critical points.}
\end{figure}

Based on the approximate free energy of Eqs.~(\ref{EQfid})-(\ref{EQfhs}), we 
have performed a coexistence analysis to determine the equilibrium fluid-fluid 
demixing binodal, defined by equality of pressures and chemical potentials 
in coexisting colloid-rich and colloid-poor fluid phases.  We restrict 
our study to parameters for which crystallization occurs only at 
higher colloid densities, well separated from fluid-fluid demixing.
Figure~\ref{FIGpd} presents the resulting phase diagrams for a fixed
polymer-to-colloid size ratio, $q=1$, and for effective polymer-to-colloid 
size ratios $\q=1$, $0.8$, $0.6$, $0.4$, corresponding to 
electrostatic colloid size ratios $\xi=1$, $1.11$, $1.25$, $1.43$, 
{\it i.e.}, varying ionic strengths [varying $\kappa$ in Eq.~(\ref{EQvcc})].
We assume, as discussed in Sec.~\ref{SECtheory}, that $\xi$ is independent of 
colloid density, which is valid in the high-salt-concentration regime 
($\rho_s\gg Z\rho_c$), where salt ions overwhelm counterions and $\kappa$ 
[Eq.~(\ref{kappa})] is essentially independent of $\rho_c$.
In the extreme limit, $\kappa\to\infty$ ($\xi\to 1$), we recover the binodal 
for the purely entropic case of neutral colloids.  With decreasing ionic 
strength (decreasing $\kappa$, increasing effective colloid radius), the 
binodal shifts to significantly higher polymer concentrations 
and the critical point to slighty lower colloid concentrations. 
The qualitative consequence of lowering ionic strength is thus a 
significant enhancement of stability against fluid-fluid demixing.
In the absence of polymer ($\eta_p=0$), the mixture reduces to a 
one-component system of hard spheres of effective radius $\Rc$.
In this limit, crystallization occurs at effective 
colloid volume fraction $\etac=\xi^3\eta_c=0.494$, which is well above 
the demixing critical volume fractions for $\xi$ values considered here
(see Fig.~\ref{FIGpd}). 
Note that the case $\xi=1.43$ is close to the limit beyond which 
demixing becomes metastable with respect to the freezing transition. 
At higher values of $\xi$, freezing may preempt demixing -- a scenario 
that we do not explore here.

The enhanced stability of charged-colloid -- neutral-polymer mixtures 
can be physically interpreted in terms of a weakening of polymer 
depletion-induced colloidal attraction.  Electrostatic repulsion 
increases the average separation between macroions, tending to lower
the frequency of configurations in which polymer is depleted from the 
intervening space.  In our model, as the range of electrostatic repulsion 
increases ({\it e.g.}, by removing salt), the effective polymer-to-colloid 
size ratio decreases, reducing the range of the effective colloid-colloid 
attraction, and thus diminishing the driving force for phase separation.

\begin{figure}
\includegraphics[width=0.9\columnwidth,angle=0]{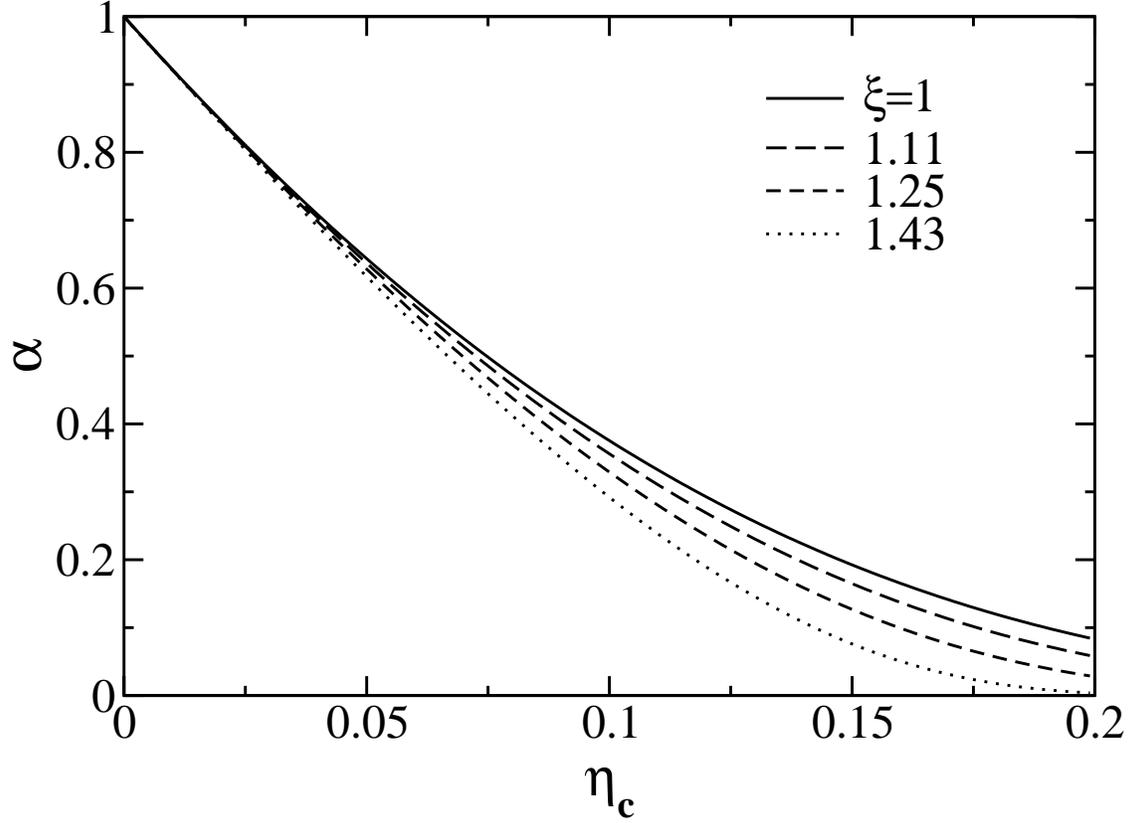}
\caption{\label{FIGalpha}Polymer free-volume fraction $\alpha$ 
[from Eq.~(\ref{EQalpha})] as a function of colloid volume fraction $\eta_c$ 
for fixed polymer-to-colloid size ratio, $q=R_p/R_c=1$, 
and varying electrostatic colloid size ratio $\xi=\Rc/R_c$.  
From top to bottom, $\xi=1, 1.11, 1.25, 1.43$, corresponding to 
increasing effective colloid radius.}
\end{figure}

A  physically equivalent interpretation of the electrostatic suppression 
of demixing follows from considering the polymer free-volume fraction 
$\alpha$.  From inspection of Eq.~(\ref{EQalpha}), it is not immediately 
obvious how $\alpha$ varies with electrostatic colloid size ratio $\xi$, 
given that, with increasing $\xi$, the quantities $(1-\etac)$, $A$, $B$, 
and $C$ all decrease, while $x$ increases.  As Fig.~\ref{FIGalpha} illustrates, 
however, with increasing electrostatic repulsion between colloids, the 
free volume available to the polymer coils is monotonically reduced, 
even though the colloid-polymer interaction is unchanged.
The reason for this reduction is that more strongly repelling colloids 
remain more widely separated, allowing less excluded volume to be hidden 
within overlapping polymer depletion shells surrounding the colloids.  
The lower free-volume fraction reduces the polymer entropy in the
colloid-rich phase, thereby suppressing demixing.

\begin{figure}
\hspace{-2cm}
\includegraphics[width=\mypicwidth,angle=-90]{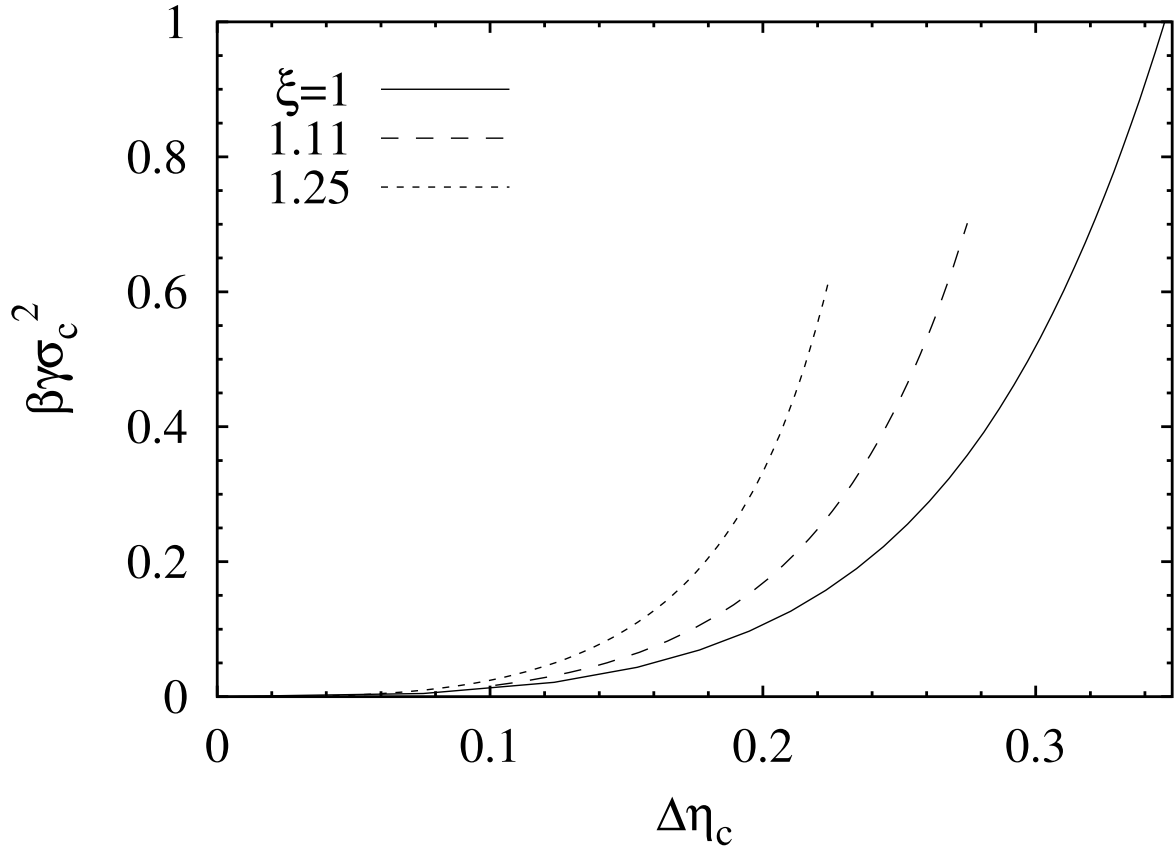}
\caption{\label{FIGpgamma2}Interfacial tension $\gamma$, in reduced units, 
between liquid (colloid-rich) and vapor (colloid-poor) phases vs. the 
difference in colloid volume fraction $\Delta\eta_c$
between colloidal liquid and vapor (at bulk coexistence).}
\end{figure}

As a check on the consistency of the above interpretations, we consider
the liquid-vapor interfacial tension $\gamma$ as a function of the 
colloid volume fraction difference $\Delta\eta_c$ between the coexisting
phases.  To calculate $\gamma$, we exploit previous density-functional (DF)
theory results~\cite{brader02,wessels04} for the AO model, which have
been found to be in reasonable agreement with 
simulation~\cite{vink04jcp,vink04codef}.
By appropriately rescaling the DF predictions of Ref.~\onlinecite{brader02} 
($\Delta\eta_c\to\Delta\eta_c/\xi^3$ and $\beta\gamma\sigma_c^2\to\beta
\gamma\sigma_c^2/\xi^2$), we obtain the results shown in Fig.~\ref{FIGpgamma2}.
Evidently, the interfacial tension {\em increases} with increasing 
effective colloid radius, which is consistent with the predicted suppression 
of liquid-vapor demixing.
Although the mapping of DF theory from neutral to charged systems is 
straightforward in the absence of external potentials, more care may be 
required in the presence of a surface.  For example, a hard wall for 
the neutral system becomes a non-additive wall for the charged system.  
In such cases, coupling to the external potential may have to be 
explicitly built into the theory.

\section{Summary and Conclusions}
\label{SECconclusions}
In summary, we have investigated fluid-fluid demixing in mixtures of 
charged colloids and neutral non-adsorbing polymers.  Our theoretical
approach involves first mapping a model mixture of charged hard-sphere 
colloids -- interacting via an effective Yukawa electrostatic pair potential
-- and ideal polymer spheres onto an effective Asakura-Oosawa model with 
purely excluded-volume interactions, and then applying free-volume theory
and density-functional theory to study, respectively, bulk phase separation 
and interfacial tension.  The effective model is characterized by a
single dimensionless parameter, namely the ratio of the polymer diameter 
to the effective colloid diameter. 

Within this simple framework, we find that increasing the range of the 
electrostatic repulsion between macroions ({\it e.g.}, by decreasing
salt concentration) suppresses demixing and increases the interfacial 
tension between coexisting colloid-rich and colloid-poor phases.
Hence, our main conclusion is that electrostatic interactions stabilize 
the suspension against depletion-induced phase separation.  As the 
depletion-induced attraction is short-ranged, the principal mechanism 
is analogous to charge-stabilization against coagulation induced by 
van der Waals forces.  In both cases the electrostatic repulsion keeps 
the colloidal particles (at least partially) outside the range of attraction.
Within our free-volume approach, the effect is clearly manifested
through a {\rm reduction} of the effective polymer size, and a
correspondingly weaker depletion-induced attraction.
The predicted increase of miscibility with decreasing salt concentration is 
qualitatively consistent with the observed behavior of mixtures of charged 
proteins and uncharged non-adsorbing polysaccharides ({\it e.g.}, gelatin 
and dextran), which are constituents of many food colloids~\cite{tolstoguzov91}.
Future experiments and simulations would help to further test the 
qualitative trends predicted here.

A possible extension of the present theory, which could yield more 
quantitative predictions, would be based on a variational method for 
the free energy.  This method would split the free energy into two parts: 
a (zeroth order) reference term, describing the effective AO model, 
and a (first-order) perturbative term, describing the repulsive tail 
of the colloidal pair potential.  Minimizing the total free energy 
with respect to the effective colloid hard-sphere diameter would give 
an upper bound on the true free energy.  Such an approach might be 
better suited to describing colloid-polymer mixtures with long-range 
electrostatic repulsion between the colloids, which have been studied 
in recent experiments~\cite{schurtenberger04,bartlett04}, and could be 
applied also to mixtures of colloids and nonideal polymers, {\it e.g.}, 
polyelectrolytes~\cite{mondain-monval95,walz, petekidis02,helden04,ferreira00},
properly incorporating soft polymer-polymer interactions~\cite{louis00}.

In this paper, we have focused attention on bulk fluid-fluid demixing in 
the high-salt-concentration regime.  As an outlook for future work, we 
mention two related phenomena that could be addressed within a similar 
theoretical framework.  
First, the fluid-solid (freezing) phase behavior of colloid-polymer mixtures
is likely to be enriched by combining electrostatic and polymer-depletion 
interactions.  It is well known~\cite{pusey91,monovoukas89} that concentrated 
suspensions of charged colloids crystallize into close-packed fcc or hcp 
structures at higher ionic strengths, and into the more open bcc structure 
at lower ionic strengths.  It would be interesting to examine the competing 
influences of electrostatics and polymer depletion on freezing and the 
relative stabilities of crystal structures.

Second, strongly deionized suspensions of highly charged colloids are known
to exhibit surprisingly complex phase behavior.
For example, experimental observations~\cite{tata92,ise94} 
and theoretical predictions~\cite{vanroij97,vanroij99,warren00} indicate
the possibility of unusual bulk separation into macroion-rich and -poor 
phases, driven by a competition between counterion entropy and 
macroion-counterion attractive energy. 
Incorporating the volume energy [Eq.~(\ref{evol})] into the total free energy
would make possible an exploration of the influence of added depletants 
on counterion-facilitated phase separation in deionized suspensions.

\begin{acknowledgments} 
This work was supported by the National Science Foundation under
Grant No.~DMR-0204020 and by the SFB TR6 of the DFG.  
Helpful discussions with Moreno Fasolo and Remco Tuinier are 
gratefully acknowledged.
\end{acknowledgments} 




\newpage




\unitlength1mm

\end{document}